\documentclass[aip,jcp,reprint,nofootinbib]{revtex4-2}
\usepackage{graphicx,amssymb,amsmath,mhchem}
\usepackage{hyperref}
\usepackage{color}
\usepackage[dvipsnames]{xcolor}
\begin{document}
\title{Negative Differential Capacitance from Composition-Dependent Stern Capacitance in a Binary Mixture}
\author{Yuki Uematsu}
\email{uematsu.yuki.2v@kyoto-u.ac.jp}
\affiliation{Department of Physics, Kyoto University, Kyoto  606-8502, Japan}
\date{\today}

\begin{abstract}
We develop a thermodynamic theory of electric double layers in binary liquid mixtures by allowing the Stern-layer capacitance to depend on the local solvent composition.
This coupling produces an additional negative contribution to the inverse differential capacitance.
As a result, the surface potential can become a nonmonotonic function of the surface charge density, leading to negative differential capacitance and a voltage-induced first-order transition between two electric-double-layer states.
We determine the coexistence condition using a common-tangent construction for the surface-charge-controlled grand potential and obtain phase diagrams in terms of the Stern-capacitance contrast, surface charge density, bulk composition, and surface potential.
We also compare the theory with capacitance data for tetrabutylammonium chloride in water/1-propanol mixtures, finding semi-quantitative agreement in the continuous-response regime.
These results suggest that solvent exchange inside the Stern layer can strongly control the capacitance and interfacial phase behavior of binary mixtures.
\end{abstract}

\maketitle

\section{Introduction}

The theory of the electric double layer was developed by Gouy, Chapman, and Stern over one hundred years ago \cite{Gouy_1910, Chapman_1913, Stern_1924}. 
Since then, their work has significantly contributed to colloid and interface science, electrochemistry, and soft matter physics. 
In the Gouy-Chapman theory, thermal diffusion and electrostatic interactions between a charged surface and ions are treated at the mean-field level by using uniform electrochemical potentials of the ions \cite{Gouy_1910,Chapman_1913}. 
Stern then proposed a thin insulating layer near the interface, which suppresses the differential capacitance through a series capacitance \cite{Stern_1924}.
The theory explains the shape of electrocapillary curves and the differential capacitance of interfaces between electrolytes and metals \cite{Grahame_1947,Valette_1981}.
Modern theories of the electric double layer are based on the Gouy-Chapman-Stern theory \cite{Kilic_2007, Uematsu_2018, Becker_2023}. 

Once the solvent of an electrolyte solution is replaced by a binary mixture, it is necessary to consider additional effects in the double-layer theory \cite{Ben_Yaakov_2009, Yabunaka_2017, Uematsu_2025}. 
One of the important effects is the composition-dependent dielectric constant.
When the binary mixture consists of water (high dielectric liquid, $\varepsilon \approx 80$) and a polar organic liquid (low dielectric liquid, $\varepsilon \approx 20$), weak water adsorption is induced inside the diffuse layer \cite{Ben_Yaakov_2009,Uematsu_2025}, because the electric field thermodynamically favors the high-dielectric liquid.
Furthermore, when the preferential solvation of ions by each solvent component is taken into account, a voltage-induced first-order surface phase transition is predicted by Yabunaka et al. \cite{Yabunaka_2017}.
In their theory, antagonistic salts, which have a strong asymmetry in the preferential solvation of cations and anions, play a key role in the phase transition and coexistence of the electric double layers.  
Furthermore, analytical results for the capacitance of the double layer at the point of zero charge were derived by Uematsu \cite{Uematsu_2025}.
When the binary mixture is sufficiently close to the bulk instability line of phase separation, the capacitance at the point of zero charge diverges and strongly deviates from the classical Gouy-Chapman result within the single-liquid approximation \cite{Uematsu_2025}.
Although these theoretical studies revealed novel properties of the electric double layers in binary mixtures, experimental validation and agreement are still lacking. 
One possible reason for this mismatch is that these theories consider only the diffuse layer, without including the Stern layer.    
Experimental measurements on the double-layer capacitance of interfaces between polycrystalline metal electrodes and binary mixtures have revealed that the capacitance is governed by the Stern-layer capacitance, rather than the diffuse-layer capacitance \cite{Aoki_2018, Iwasaki_2023}. 
Furthermore, they observed strong water adsorption to the electrodes even when the water fraction is very low in the bulk \cite{Aoki_2018, Iwasaki_2023}.

Thus, the motivation of this paper is to study the double-layer capacitance of binary mixtures by including the effect of the Stern layer over a wide voltage range.  
The main difference between the Stern layer in a single liquid and that in a binary mixture is that the solvent composition in the Stern layer can differ from both the bulk composition and the composition at the outer edge of the diffuse layer, owing to short-range interactions between the surface and the solvent molecules. 
This composition-dependent Stern-layer capacitance can induce an additional contribution to the total capacitance beyond the direct series combination of a constant Stern capacitance and the diffuse-layer capacitance.  
We formulate the thermodynamic model to include the composition-dependent Stern-layer capacitance. 
Interestingly, we find that this effect always contributes to the instability of the differential capacitance in voltage-controlled systems.
As a result, we predict a voltage-induced surface phase transition and coexistence even without preferential solvation or the bulk instability of phase separation, which were reported previously \cite{Yabunaka_2017, Uematsu_2025}. 
The critical condition for the contrast between the Stern-layer capacitances of the two pure liquids is derived and discussed in relation to previously obtained experimental data. 
Furthermore, we also compare our theory with recent experimental observations on the capacitance as a function of the liquid composition. 
We believe that these findings and discussions provide insight into a wide range of problems in colloid and interface science, such as colloids, electrokinetics, and electrochemistry in binary mixtures.

\section{Model}
Although the model described in this section is similar to those in Refs.~\citenum{Ben_Yaakov_2009, Yabunaka_2017, Uematsu_2025}, the key novelty is that we consider the Stern layer explicitly in the free-energy formulation. 
The Helmholtz free energy of an electric double layer in a binary mixture per unit area is given by
\begin{equation}
F = F_\mathrm{s} + F_\mathrm{d},
\end{equation}
where $F_\mathrm{s}$ is the free energy of the Stern layer and $F_\mathrm{d}$ is the free energy of the diffuse layer.  
The Stern layer has a thickness of $z_\mathrm{s}$, and thus, the free energy of the diffuse layer is 
\begin{equation}
F_\mathrm{d}  = \int^\infty_{z_\mathrm{s}}  \left[f(\phi,c_+,c_-)+f_\mathrm{g}+f_\mathrm{e} - f_\infty\right] dz,
\end{equation}
where $f(\phi,c_+,c_-)$ is the chemical part of the Helmholtz free energy density, $\phi$ is the local volume fraction of the second liquid (organic fraction), $c_+$ and $c_-$ are the local concentrations of the cations and anions, $f_\mathrm{g}$ is the gradient free energy, and $f_\mathrm{e}$ is the electrostatic energy density.
The last term in the integral, $f_\infty = f(\phi_\mathrm{b},c_\mathrm{b}, c_\mathrm{b})$, is a constant that ensures convergence of the integral, and $\phi_\mathrm{b}$ and $c_\mathrm{b}$ are the bulk organic fraction and ion concentration.  
We assume an isothermal and isochoric process. 
Furthermore, we consider the situation where the volume fraction of the ions is negligible. 
Assuming the first and second solvents have equal molecular volume, $v_0$, the Helmholtz free energy density at the mean-field level is
\begin{equation}
\begin{split}
& f(\phi,c_+,c_-)  = k_\mathrm{B}T\sum_{i=\pm}c_i\left[\ln(c_i v_0)-1+\alpha_i\phi\right]\\
& + \frac{k_\mathrm{B}T}{v_0}\left[\phi\ln\phi+(1-\phi)\ln(1-\phi)+\chi\phi(1-\phi)\right], 
\end{split}
\label{eq:2}
\end{equation}
where $k_\mathrm{B}$ is the Boltzmann constant, $T$ is the temperature, $\alpha_i$ are the parameters for preferential solvation of ions, and $\chi$ is the parameter for the interaction between two different solvent molecules.  
The gradient free energy is given by \cite{Yabunaka_2017}
\begin{equation}
f_\mathrm{g} = \frac{K}{2}(\nabla\phi)^2,
\end{equation}
where $K$ is a constant of order $k_\mathrm{B}T/{v_0}^{1/3}$.
The electrostatic energy density is given by
\begin{equation}
f_\mathrm{e}=\frac{\varepsilon(\phi)\varepsilon_0}{2}(\nabla\psi)^2,
\end{equation}
where $\varepsilon(\phi)$ is the dielectric constant, $\varepsilon_0$ is the electric permittivity of vacuum, and $\psi$ is the local electrostatic potential.  
The dependence of the dielectric constant on the organic fraction is considered to be linear, as given by
\begin{equation}
\varepsilon(\phi) = \varepsilon_1+(\varepsilon_2-\varepsilon_1)\phi,
\end{equation}
where $\varepsilon_1$ and $\varepsilon_2$ are the dielectric constants of the first and second solvents.
We assume that the first solvent has a high dielectric constant like water, whereas the second solvent is polar but has a lower dielectric constant. 
For concreteness, we hereafter refer to the first and second liquid components as water and alcohol, respectively, although the formulation applies to a general binary liquid mixture.

Regarding the free energy of the Stern layer, ion adsorption at the interface between the Stern and diffuse layers is often considered using the Langmuir adsorption isotherm \cite{Stern_1924}. 
Here, it is not included in the model for simplicity.
Furthermore, the Stern layer is considered to be insulating and does not include ions.
Therefore, the free energy of the Stern layer is composed of the electrostatic energy, the entropic contribution of the liquid component in the Stern layer, and the energy term due to the interaction between the surface and molecules, as given by 
\begin{equation}
\begin{split}
F_\mathrm{s} & = \frac{k_\mathrm{B}T z_\mathrm{s}}{v_0} \left[\phi_0\ln\phi_0+(1-\phi_0)\ln(1-\phi_0)+\alpha_0\phi_0\right]\\
& +\frac{{\sigma_0}^2}{2C_\mathrm{s}(\phi_0)},
\end{split}
\end{equation}
where $\phi_0$ is the organic fraction in the Stern layer, $\alpha_0$ is a dimensionless parameter, $\sigma_0$ is the surface charge density, and $C_\mathrm{s}$ is the Stern-layer capacitance. 
Here, the Stern-layer capacitance is assumed to be composition-dependent, as given by
\begin{equation}
\frac{1}{C_\mathrm{s}(\phi_0)} = \frac{1}{C_1} + \left(\frac{1}{C_2}-\frac{1}{C_1}\right)\phi_0 = \frac{1}{C_1}+\gamma\phi_0,
\end{equation}
where $C_1$ and $C_2$ are the Stern-layer capacitances of the first and second liquids (water and alcohol), and $\gamma$ is the Stern-capacitance contrast with the inverse unit of surface capacitance.
In the present notation, positive $\gamma$ means $C_1>C_2$, namely that the Stern capacitance decreases as the Stern layer becomes rich in the alcohol.
Because $\phi_0$ denotes the organic fraction, a positive $\alpha_0$ in $F_\mathrm{s}$ disfavors the alcohol in the Stern layer, whereas a negative $\alpha_0$ favors it.

Because the Helmholtz free energy should depend only on extensive quantities (or their densities) and temperature, it is useful to write the total electrostatic energy per area in terms of $\phi$, $c_+$, $c_-$.
Defining 
\begin{equation}
F_\mathrm{e}= \frac{{\sigma_0}^2}{2C_\mathrm{s}(\phi_0)} + \int^\infty_{z_\mathrm{s}} f_\mathrm{e} dz,
\end{equation}
it can be rewritten into
\begin{equation}
\begin{split}
F_\mathrm{e}& = \int_{z_\mathrm{s}}^\infty \left[-\frac{\varepsilon(\phi)\varepsilon_0}{2}\left(\frac{d\psi}{dz}\right)^2 + \rho\psi \right] dz + \sigma_\mathrm{s}\psi_\mathrm{s} + \frac{{\sigma_0}^2}{2C_\mathrm{s}},\\
& = \int_{z_\mathrm{s}}^\infty \left[-\frac{\varepsilon(\phi)\varepsilon_0}{2}\left(\frac{d\psi}{dz}\right)^2 + \rho\psi \right] dz - \frac{{\sigma_0}^2}{2C_\mathrm{s}}+\sigma_0\psi_0,
\label{eq:electrostatics}
\end{split}
\end{equation}
where $\rho = e(c_+-c_-)$ is the charge density in the diffuse layer, $e$ is the elementary charge, $\psi_\mathrm{s}=\psi(z_\mathrm{s})$ and $\sigma_\mathrm{s}=-\varepsilon(\phi_\mathrm{s})\varepsilon_0 d\psi/dz|_{z=z_\mathrm{s}+0}$ are the electrostatic potential and the surface charge density at the interface between the Stern and diffuse layers, and $\psi_0=\psi(0)$ is the surface potential.  
The orgnic fraction at $z=z_\mathrm{s}+0$ is defined as $\phi_\mathrm{s}$, which is generally different from $\phi_0$.
Eq.~\eqref{eq:electrostatics} can be derived from the Poisson equation given by
\begin{equation}
\frac{d}{dz}\left(\varepsilon(\phi)\varepsilon_0\frac{d\psi}{dz}\right) = -\rho, \label{eq:Poisson}
\end{equation}
and the relations $\sigma_\mathrm{s}=\sigma_0$ (no ion adsorption at the Stern layer) and $\psi_\mathrm{s}=\psi_0-\sigma_0/C_\mathrm{s}$. 
The second line in Eq.~\eqref{eq:electrostatics} is the thermodynamic function for $\rho(z)$, $\sigma_0$, $\phi(z)$, and  $\phi_0$.
The derivative form of Eq.~\eqref{eq:electrostatics} with respect to  $\delta \rho(z)$, $d\sigma_0$, $\delta \phi(z)$, and $d\phi_0$ is 
\begin{equation}
\begin{split}
dF_\mathrm{e} & = \int^\infty_{z_\mathrm{s}} \left[ -\frac{\varepsilon_0}{2}\left(\varepsilon_2-\varepsilon_1\right)\left(\frac{d\psi}{dz}\right)^2\delta\phi+\psi\delta\rho\right]dz \\
& + \left(\psi_0 -\frac{\sigma_0}{C_\mathrm{s}}\right)d\sigma_0 -\frac{\gamma{\sigma_0}^2}{2}d\phi_0.
\end{split}
\end{equation}

To obtain the equilibrium profiles of $\phi(z)$, $c_i(z)$, and the composition $\phi_0$ for a given bulk salt concentration $c_\mathrm{b}$, bulk organic fraction $\phi_\mathrm{b}$, and surface charge density $\sigma_0$, we convert the Helmholtz free energy to the surface-charge-controlled grand potential $\Omega(\sigma_0)$ by a Legendre transformation, given as
\begin{equation}
\begin{split}
& \Omega(\sigma_0) = F -\mu_\phi\phi_0 z_\mathrm{s}- \int^\infty_{z_\mathrm{s}} \mu_\phi(\phi(z)-\phi_\mathrm{b})dz\\
&  - \int^\infty_{z_\mathrm{s}} \left[\mu_+(c_+(z)-c_\mathrm{b})+\mu_-(c_-(z)-c_\mathrm{b})\right]dz,
\end{split}
\end{equation}
where $\mu_\pm$ are the chemical potentials of the cation and anion, and $\mu_\phi$ is the chemical potential for the solvent composition.
Therefore, the equilibrium conditions are given by 
\begin{eqnarray}
\frac{\delta\Omega}{\delta c_i(z)} &=& 0,\label{eq:chemion} \\
\frac{\delta\Omega}{\delta\phi(z)} &=& 0,\label{eq:chemsol} \\
\frac{\partial\Omega}{\partial\phi_\mathrm{s}} &=& 0, \label{eq:bc2} \\
\frac{\partial\Omega}{\partial\phi_0} &=& 0, \label{eq:bc4}
\end{eqnarray}
where Eqs.~\eqref{eq:chemion} and \eqref{eq:chemsol} are the homogeneous conditions for the chemical potentials of the solvents and the ions in the diffuse layer.
Eq.~\eqref{eq:chemion} yields the concentrations of cations and anions as
\begin{eqnarray}
c_+(z) & = & c_\mathrm{b}\mathrm{e}^{-\Psi(z)-\alpha_+(\phi(z)-\phi_\mathrm{b})},\label{eq:boltzmann1}\\ 
c_-(z) & = & c_\mathrm{b}\mathrm{e}^{ \Psi(z)-\alpha_-(\phi(z)-\phi_\mathrm{b})},\label{eq:boltzmann2}
\end{eqnarray}
where $\Psi= e\psi/k_\mathrm{B}T$ is the dimensionless potential. 
Eq.~\eqref{eq:Poisson} with Eqs.~\eqref{eq:boltzmann1} and \eqref{eq:boltzmann2} yields the Poisson-Boltzmann equation in a binary mixture, as given by
\begin{equation}
\begin{split}
& \frac{d}{dz}\left[\varepsilon(\phi)\varepsilon_0\frac{d\psi}{dz}\right] \\
& = -ec_\mathrm{b}\left[\mathrm{e}^{-\Psi(z)-\alpha_+(\phi(z)-\phi_\mathrm{b})}-\mathrm{e}^{ \Psi(z)-\alpha_-(\phi(z)-\phi_\mathrm{b})}\right].
\end{split}
\label{eq:PB}
\end{equation}
Eqs.~\eqref{eq:chemsol} and \eqref{eq:bc2} give the conditions for the uniform chemical potential of the solvent in the diffuse layer, given as 
\begin{equation}
\begin{split}
&-\frac{(\varepsilon_2-\varepsilon_1)\varepsilon_0}{2k_\mathrm{B}T}\left(\frac{d\psi}{dz}\right)^2 +\alpha_+(c_+-c_\mathrm{b})+\alpha_-(c_--c_\mathrm{b})\\
&+\frac{1}{v_0}\left[\ln\frac{\phi}{1-\phi}-\ln\frac{\phi_\mathrm{b}}{1-\phi_\mathrm{b}}-2\chi(\phi-\phi_\mathrm{b})\right] \\
& -\frac{K}{k_\mathrm{B}T}\frac{d^2\phi}{dz^2}=0,
\end{split}
\label{eq:g}
\end{equation}
with the boundary condition 
\begin{equation}
\left.\frac{d\phi}{dz}\right|_{z=z_\mathrm{s}}=0.
\label{eq:bcphi}
\end{equation}
The remaining boundary conditions used for the diffuse layer are
\begin{equation}
-\varepsilon(\phi_\mathrm{s})\varepsilon_0
\left.\frac{d\psi}{dz}\right|_{z=z_\mathrm{s}+0}
=\sigma_0,\quad
\psi(\infty)=0,\quad
\phi(\infty)=\phi_\mathrm{b}.
\end{equation}
Eq.~\eqref{eq:bc4} gives the adsorption isotherm of the solvents as
\begin{equation}
\phi_0 = \frac{\phi_\mathrm{b}\mathrm{e}^{\chi(1-2\phi_\mathrm{b})+(\alpha_++\alpha_-)c_\mathrm{b}v_0-\alpha_0-\gamma{\sigma_0}^2v_0/2k_\mathrm{B}Tz_\mathrm{s}}}{1-\phi_\mathrm{b}+\phi_\mathrm{b}\mathrm{e}^{\chi(1-2\phi_\mathrm{b})+(\alpha_++\alpha_-)c_\mathrm{b}v_0-\alpha_0-\gamma{\sigma_0}^2v_0/2k_\mathrm{B}Tz_\mathrm{s}}}.
\label{eq:phi0}
\end{equation}
In the numerical calculations below, these equations are solved as a boundary-value problem at fixed $\sigma_0$, and $\psi_0$ is then obtained from $\psi_0=\psi_\mathrm{s}+\sigma_0/C_\mathrm{s}(\phi_0)$.

\section{Results and Discussion}

\begin{figure}
\center
\includegraphics[width=8.5cm]{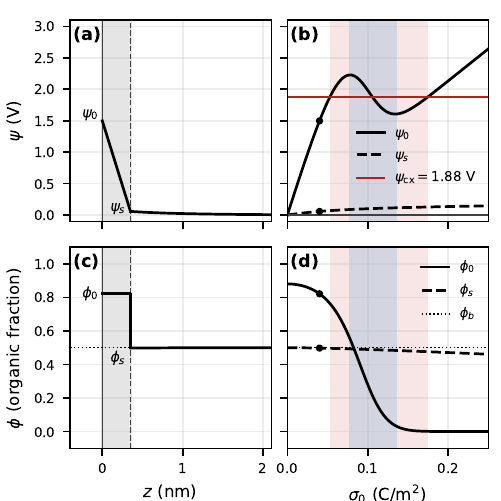}
\caption{
(a) Profile of the electrostatic potential at $\sigma_0=0.04\,$C/m$^2$.
(b) Surface and interfacial potentials as a function of $\sigma_0$. The red line denotes the coexistence potential. 
(c) Profile of the organic fraction at $\sigma_0=0.04\,$C/m$^2$.
(d) Bulk, interfacial and Stern-layer organic fractions as a function of $\sigma_0$.
The black points in (b) and (d) denote the states at $\sigma_0 = 0.04\,$C/m$^2$.
The blue region denotes negative differential capacitance, whereas the red region denotes the binodal region. 
}
\label{fig:1}
\end{figure}

We numerically solve Eqs.~\eqref{eq:PB}, \eqref{eq:g}, \eqref{eq:bcphi}, and \eqref{eq:phi0} and calculate $\psi(z)$ and $\phi(z)$.
Fig.~\ref{fig:1}(a) and (c) show typical results of the profiles at $\sigma_0=0.04\,$C/m$^2$, whereas (b) and (d) show $\psi_0$, $\psi_\mathrm{s}$, $\phi_0$, and $\phi_\mathrm{s}$ as a function of $\sigma_0$.   
The parameters used are $\sigma_0=0.04\,$C/m$^2$, $\phi_\mathrm{b}=0.5$, $c_\mathrm{b}=0.1\,$M, $\alpha_+=\alpha_-=\chi=0$, $\varepsilon_1=80$, $\varepsilon_2=20$, $v_0=27\,$\AA$^3$, $C_1=0.1\,$F/m$^2$, $C_2=0.024\,$F/m$^2$, $\alpha_0=-2$, $z_\mathrm{s}=0.35\,$nm, and $T = 300\,$K.   
The profile of $\psi(z)$ is similar to that in the Gouy-Chapman-Stern theory for a single liquid. 
The potential decays linearly in the Stern layer and exponentially in the diffuse layer.
On the other hand, $\phi_0=0.83$ shows that the Stern layer is alcohol-rich, reflecting the non-electrostatic adsorption energy $\alpha_0=-2$, whereas the bulk organic fraction is $\phi_\mathrm{b}=0.5$. 
$\phi_\mathrm{s}=0.498$ is slightly lower than $\phi_\mathrm{b}$, suggesting weak water adsorption \cite{Ben_Yaakov_2009, Uematsu_2025}.
As shown in Fig.~\ref{fig:1}(b), when the surface charge density increases, the surface potential starts to decrease at $\sigma_0=0.078\,$C/m$^2$ and then increases again at $\sigma_0=0.134\,$C/m$^2$. 
This curve suggests negative differential capacitance even in the absence of preferential solvation ($\alpha_+=\alpha_-=0$) and bulk instability of phase separation ($\chi=0$).
In the range of negative differential capacitance, $\psi_\mathrm{s}$ is still a monotonically increasing function of $\sigma_0$, whereas $\phi_0$ and $\phi_\mathrm{s}$ decrease monotonically, as shown in Fig.~\ref{fig:1}(d). 

To see how the negative differential capacitance is induced, we calculate the inverse of the differential capacitance analytically. 
The total differential capacitance of the electric double layer is defined as $C=d\sigma_0/d\psi_0$ and is calculated by 
\begin{equation}
\begin{split}
\frac{1}{C} & = \frac{d\psi_0}{d\sigma_0} = \frac{d\psi_\mathrm{s}}{d\sigma_0}+\frac{1}{C_\mathrm{s}}+\gamma\sigma_0\frac{d\phi_0}{d\sigma_0}\\
& = \frac{d\psi_\mathrm{s}}{d\sigma_\mathrm{s}}+\frac{1}{C_\mathrm{s}}- \frac{\gamma^2{\sigma_0}^2v_0}{k_\mathrm{B}T z_\mathrm{s}}\phi_0(1-\phi_0),
\end{split}
\label{eq:total_cap}
\end{equation}
using the differentiation of Eq.~\eqref{eq:phi0}. 
The first term $d\psi_\mathrm{s}/d\sigma_\mathrm{s}$ is the inverse capacitance of the diffuse layer. 
Eq.~\eqref{eq:total_cap} is different from the well-known equation for capacitances in series because of the third term.
The third term is negative except in the case of $\gamma=0$ or $\sigma_0=0$ (the point of zero charge). 

\begin{figure}
\includegraphics[width=8.5cm]{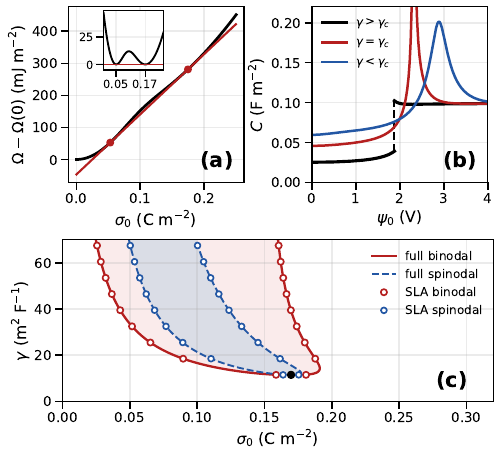}
\caption{
(a) Surface-charge-controlled grand potential relative to $\Omega(0)$. The red line denotes a common tangent. 
The inset shows the difference between $\Omega(\sigma_0)-\Omega(0)$ and the common tangent. $0.05\,$C/m$^2$ and $0.17\,$C/m$^2$ denote the coexisting states.  
(b) Differential capacitance $C= d\sigma_0/d\psi_0$ as a function of the surface potential $\psi_0$.
The dashed vertical segments indicate the discontinuous jumps at the coexistence potential.
The black line denotes $\gamma>\gamma_\mathrm{c}$, whereas the red line denotes $\gamma=\gamma_\mathrm{c}=11.5\,$m$^2$/F and the blue line denotes $\gamma=\gamma_\mathrm{c}/2$.
(c) Phase diagram of the electric double layers in the $\sigma_0$-$\gamma$ plane. 
The red solid and blue dashed lines denote the binodal and spinodal lines, respectively, and the black point marks the critical point. 
Red and blue open points are calculated using the single-liquid approximation in the diffuse layer.
We fix $C_1$ and vary $C_2$ to adjust $\gamma$ in (b) and (c).
Other parameters used are the same as those in Fig.~\ref{fig:1}. 
}
\label{fig:2}
\end{figure}

The nonmonotonic dependence of $\psi_0$ as a function of $\sigma_0$ indicates coexistence of two surface charge densities, $\sigma_0=\sigma_1$ and $\sigma_2$, with a single $\psi_0 = \psi_\mathrm{cx}$. 
Phase coexistence is obtained by constructing a common tangent to $\Omega(\sigma_0)$ as shown in Fig.~\ref{fig:2}a.
The slope of the tangent corresponds to the electrostatic potential at the transition $\psi_\mathrm{cx}$, which is denoted by the red line in Fig.~\ref{fig:1}b.
The inset of Fig.~\ref{fig:2}a shows $\Omega-\Omega_\mathrm{tan}$, with $\Omega_\mathrm{tan}=\Omega(\sigma_1)+\psi_\mathrm{cx}(\sigma_0-\sigma_1)$, making the two coexisting surface charges and the intervening barrier explicit.
Because $\psi_0 = d\Omega/d\sigma_0$, the classical Maxwell rule holds as 
\begin{equation}
\Omega(\sigma_2) - \Omega(\sigma_1) = \int^{\sigma_2}_{\sigma_1} \psi_0(\sigma_0) d\sigma_0 = \psi_\mathrm{cx}(\sigma_2-\sigma_1). 
\label{eq:binodal}
\end{equation}
Thus, the areas of the two regions in Fig.~\ref{fig:1}b split by the red line are equal.
Fig.~\ref{fig:2}b shows the differential capacitance as a function of the surface potential.
The black line denotes the capacitance with $\gamma = 31.7\,$m$^2$/F $(>\gamma_\mathrm{c}=11.5\,$m$^2$/F$)$, and it is discontinuous at $\psi_\mathrm{cx}$.
The Stern layer at low surface charge is mainly composed of the organic liquid, whereas after the transition it is almost entirely water-rich.  
Fig.~\ref{fig:2}c shows a phase diagram in the $\gamma$-$\sigma_0$ plane. 
We fix $C_1$ and vary $C_2$ to adjust $\gamma$ in (b) and (c).
Other parameters used are the same as those in Fig.~\ref{fig:1}. 
The binodal line (red line) is determined by Eq.~\eqref{eq:binodal} and the spinodal line (broken blue line) is determined by $1/C=0$. 
They end at a critical point at $(\sigma_\mathrm{c},\gamma_\mathrm{c})$ (black point), and thus, the voltage-induced transition happens only for $\gamma>\gamma_\mathrm{c}$.
In other words, no surface transition is induced by applying voltage when $\gamma<\gamma_\mathrm{c}$, as shown by the blue line in Fig.~\ref{fig:2}b.
When $\gamma=\gamma_\mathrm{c}$, the differential capacitance diverges, as shown by the red line in Fig.~\ref{fig:2}b. 

To compute the capacitance in a wide parameter range, we use the single-liquid approximation in the diffuse layer.
In this approximation, the oragnic fraction profile for $z>z_\mathrm{s}$ is approximated to be uniform and equal to $\phi_\mathrm{b}$. 
This approximation is valid at least in the absence of preferential solvation, $\alpha_+=\alpha_-=0$, and for $\chi=0$, far away from the bulk instability \cite{Uematsu_2025}.
Then, the potential profile in the diffuse layer is the Gouy-Chapman solution with the uniform dielectric constant $\varepsilon(\phi_\mathrm{b})$, and its differential capacitance is given by
\begin{equation}
\begin{split}
\frac{d\sigma_\mathrm{s}}{d\psi_\mathrm{s}} & = \varepsilon(\phi_\mathrm{b})\varepsilon_0 \sqrt{\frac{2e^2c_\mathrm{b}}{\varepsilon(\phi_\mathrm{b})\varepsilon_0 k_\mathrm{B}T}}\cosh\left(\frac{e\psi_\mathrm{s}}{2k_\mathrm{B}T}\right)\\
& =\varepsilon(\phi_\mathrm{b})\varepsilon_0 \sqrt{\frac{2e^2c_\mathrm{b}}{\varepsilon(\phi_\mathrm{b})\varepsilon_0 k_\mathrm{B}T}} \sqrt{1+\frac{{\sigma_\mathrm{s}}^2}{2\varepsilon(\phi_\mathrm{b})\varepsilon_0 k_\mathrm{B}Tc_\mathrm{b}}}.
\end{split}
\label{eq:SLA}
\end{equation}
Using Eqs.~\eqref{eq:total_cap} and \eqref{eq:SLA}, the differential capacitance is obtained without solving the coupled partial differential equations, Eqs.~\eqref{eq:PB} and \eqref{eq:g}.
The open points in Fig.~\ref{fig:2}c are calculated from the single-liquid approximation in the diffuse layer, and the agreement with the full numerical calculation is excellent.

\begin{figure}
\includegraphics[width=85mm]{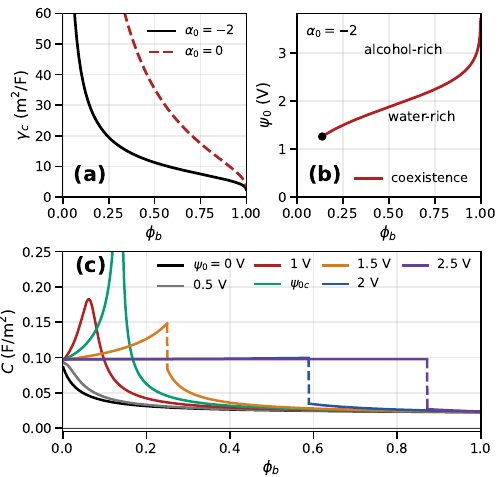}
\caption{
(a) $\gamma_\mathrm{c}$ as a function of $\phi_\mathrm{b}$. 
(b) Phase diagram in $\phi_\mathrm{b}$-$\psi_0$ plane for $\alpha_0=-2$.
(c) Differential capacitance as a function of $\phi_\mathrm{b}$.
Panels (b) and (c) are calculated at $\gamma=31.7\,$m$^2$/F and $\alpha_0=-2$.
The broken lines denote discontinuous phase transitions. 
}
\label{fig:5}
\end{figure}

Because $\gamma$ is a material parameter, it is not easy to control in experiments. 
A more easily controlled parameter is the bulk organic fraction. 
Since the single-liquid approximation in the diffuse layer shows excellent agreement with the full numerical calculation, we use it throughout the remainder of this paper to analyze the dependence on the bulk composition $\phi_\mathrm{b}$.
To understand the mechanism of the surface phase transition, $\gamma_\mathrm{c}$ is plotted in Fig.~\ref{fig:5}a as a function of $\phi_\mathrm{b}$ with $\alpha_0=-2$ and $0$. 
Here, unless otherwise stated, we use $c_\mathrm{b}=0.1\,$M, $\alpha_+=\alpha_-=\chi=0$, $\varepsilon_1=80$, $\varepsilon_2=20$, $v_0=27\,$\AA$^3$, $C_1=0.1\,$F/m$^2$, $z_\mathrm{s}=0.35\,$nm, and $T=300\,$K.
The obtained $\gamma_\mathrm{c}$ exhibits a monotonic decrease with $\phi_\mathrm{b}$ for any $\alpha_0$, suggesting asymmetry in the organic fraction of the solvent.
This asymmetry originates from the sign of $\gamma$. 
Positive $\gamma$ means that the electric field in the Stern layer favors water (the first solvent) over alcohol (the second solvent).
When the organic fraction in the Stern layer is large ($\phi_0$ is large), applying voltage induces the replacement of alcohol by water via a phase transition.
Thus, high $\phi_\mathrm{b}$ and negative $\alpha_0$ make $\gamma_\mathrm{c}$ smaller. 
By contrast, when water is dominant in the Stern layer ($\phi_\mathrm{b}\ll 1$ or $\alpha_0>0$), $\gamma_\mathrm{c}$ becomes larger.  

In Fig.~\ref{fig:5}b, the phase diagram in the $\phi_\mathrm{b}$-$\psi_0$ plane is plotted for $\alpha_0=-2$.
The critical point corresponds to $(\phi_\mathrm{bc},\psi_{0\mathrm{c}})=(0.137,1.26\,$V$)$.
For this panel and for Fig.~\ref{fig:5}c, $\gamma$ is fixed at $31.7\,$m$^2$/F, corresponding to $C_2=0.024\,$F/m$^2$ with $C_1=0.1\,$F/m$^2$.
From the critical point, the coexistence line separates the water-rich Stern layer and the alcohol-rich Stern layer. 
Fig.~\ref{fig:5}c shows the differential capacitance as a function of $\phi_\mathrm{b}$ with $\alpha_0=-2$. 
When $\phi_\mathrm{b}=0$ or $1$, $\phi_0=\phi_\mathrm{b}$ holds. 
Therefore, the deviations of $C$ from $C_1$ at $\phi_\mathrm{b}=0$ and from $C_2$ at $\phi_\mathrm{b}=1$ are induced by the diffuse layer capacitance.
At the potentials $\psi_0=0\,$V and $0.5\,$V, the differential capacitance decreases monotonically with increasing $\phi_\mathrm{b}$, whereas at $\psi_0 = 1.0\,$V an increase in the middle of $\phi_\mathrm{b}$ is observed. 
The increase in $C$ can be understood from Eq.~\eqref{eq:total_cap}. 
The third term is negative and $1/C$ has a minimum at about $\phi_0=0.5$.  
At the voltage $\psi_0=\psi_{0\mathrm{c}}$, the differential capacitance diverges at $\phi_\mathrm{b}=\phi_{\mathrm{bc}}$.
For $\psi_0>\psi_{0\mathrm{c}}$, the differential capacitance exhibits discontinuous transition at $\phi_\mathrm{b}>\phi_\mathrm{bc}$.    

To the best of our knowledge, experimental observations of a discontinuous transition in the differential capacitance of binary mixtures have not yet been reported.
Thus, we examine the possibility of the transition.
Hou et al. measured Stern-layer capacitance in water and various organic liquids \cite{Hou_2014}. 
In Table~\ref{table:1}, the extracted Stern-layer capacitances from this reference are listed.
Choosing the combination of water and propylene carbonate, $\gamma = 1/C_2-1/C_1 = 8.3\,$m$^2$/F is achieved.
Such a high $\gamma$ can induce the transition of the double layers when $\phi_\mathrm{b}$ is sufficiently large, as shown in Fig.~\ref{fig:5}a. 

\begin{table}
\caption{Extracted Stern-layer capacitances of water and various organic liquids in contact with a polycrystalline platinum electrode \cite{Hou_2014}.}
\label{table:1}
\begin{tabular}{lrr}
\hline
solvent 	& 	$C_1$ or $C_2$ (F/m$^2$)	& 	$\gamma$ (m$^2$/F) \\\hline
water		&	0.35		& 	--\\
formamide 	& 	0.24		& 	1.3\\
ethanol		&	0.11		& 	6.2\\
propylene carbonate & 	0.09		&	8.3\\
\hline
\end{tabular}
\end{table}

\begin{figure}
\includegraphics[width=6cm]{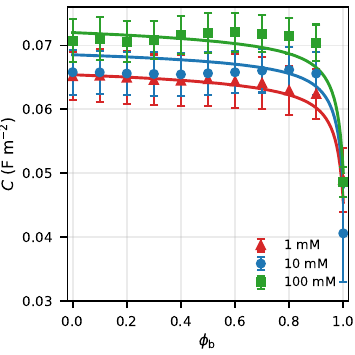}
\caption{
Comparison between the finite-potential capacitance predicted by the present model and experimental data for TBAC in water/1-propanol mixtures \cite{Iwasaki_2023}.
The horizontal axis is the bulk organic volume fraction $\phi_\mathrm{b}$.
The symbols denote the data at 1, 10, and 100 mM, and the solid curves are calculated with common fitted parameters $C_1=0.078\,$F/m$^2$, $C_2=0.055\,$F/m$^2$, $\alpha_0=4$, and $\psi_0=0.22\,$V.
}
\label{fig:4}
\end{figure}

Experimentally, the capacitances in binary mixtures show a trend consistent with strong water adsorption \cite{Aoki_2018, Iwasaki_2023}. 
Even when the organic fraction in bulk is high, the capacitance is similar to that in pure water. 
These behaviors have not been explained by the theory of the double-layer capacitance at the potential of zero charge \cite{Iwasaki_2023, Uematsu_2025}.
However, the actual electrode (surface) potential in the two-electrode measurements is not necessarily equal to the potential of zero charge.    
Thus, we fit the experimental data of capacitances in tetrabutylammonium chloride (TBAC) water/1-propanol solutions with an aluminum electrode \cite{Iwasaki_2023} using the single-liquid approximation in the diffuse layer at finite potential.
The fitted results are shown in Fig.~\ref{fig:4}. 
Because there are many parameters in the model, we fix $\alpha_+=\alpha_-=\chi=0$. 
Furthermore, we fix $\alpha_0=4$ in the comparison with the TBAC data. 
This choice corresponds to a Stern layer that is depleted of alcohol relative to the bulk.
The same values of $C_1$, $C_2$, $\alpha_0$, and $\psi_0$ are used for all three salt concentrations, while the bulk salt concentration is set to the experimental value for each data set.
Then, we fit $C_1$, $C_2$, and $\psi_0$ by calculating the capacitance at the finite surface potential using the single-liquid approximation in the diffuse layer.
The obtained parameters are $C_1=0.078\,$F/m$^2$, $C_2=0.055\,$F/m$^2$, and $\psi_0=0.22\,$V, which corresponds to $\gamma = 5.31\,$m$^2$/F.
We compute the critical point $(\phi_\mathrm{bc},\psi_{0\mathrm{c}})$ for the fitted $\gamma$, as in Fig.~\ref{fig:5}b, and obtain $\phi_\mathrm{bc}=0.9999$ and $\psi_{0\mathrm{c}}=5.4\,$V to $5.5\,$V for different salt concentrations. 
Therefore, the variations in Fig.~\ref{fig:4} are continuous because the fitted potential is below the critical potential $\psi_{0\mathrm{c}}$. 
This semi-quantitative agreement supports the central assumption of our theory, namely that the Stern-layer capacitance is controlled by the local solvent composition.

\section{Conclusion}

We have developed a theory of the electric double layer in a binary liquid mixture by explicitly coupling the Stern-layer capacitance to the local solvent composition.
The central result is that a composition-dependent Stern capacitance gives rise to an additional negative contribution to the inverse differential capacitance.
This contribution is absent in the conventional Gouy--Chapman--Stern theory with a composition-independent Stern layer.

As a consequence, the surface potential can become a nonmonotonic function of the surface charge density even in the absence of preferential solvation of ions and far from the bulk phase-separation instability.
In the potential-controlled system, the unstable branch is replaced by coexistence between two electric double-layer states with different surface charge densities and different Stern-layer compositions.
We formulated this transition in terms of the grand potential \(\Omega(\sigma_0)\) and showed that the coexistence condition is determined by a common-tangent, or Maxwell, construction.
The resulting phase diagrams reveal a critical contrast $\gamma_\mathrm{c}$ in the Stern capacitance, above which a voltage-induced first-order surface transition occurs.
We also introduced the single-liquid approximation for the diffuse layer.
This approximation reproduces the full numerical results accurately for the present parameter regime and makes it possible to analyze the phase behavior over a wide range of bulk compositions and Stern-layer parameters.

Finally, we compared the finite-potential single-liquid approximation with experimental capacitance data for TBAC in water/1-propanol mixtures \cite{Iwasaki_2023}.
Although the comparison should be regarded as semi-quantitative rather than as a unique parameter extraction, the fitted curves reproduce the observed composition dependence of the capacitance with physically reasonable Stern capacitances.
For the fitted parameters, the critical point lies at much higher potential than the fitted surface potential, indicating that the available data correspond to the continuous-response regime without a phase transition.
This agreement supports the physical picture that the Stern-layer capacitance is governed by the local solvent composition.

The present mechanism suggests that binary mixtures offer a route to controlling electric double layers through solvent composition, not only through salt concentration or electrode potential.
It also points to the possibility of voltage-induced surface phase transitions and capacitance discontinuities in systems with sufficiently large contrast between the Stern capacitances of the two pure solvents.
These findings provide a basis for understanding how solvent composition influences colloidal stability and zeta potentials in binary liquid mixtures \cite{Schwer_1991, Hasan_2023}.

\begin{acknowledgments}
The author is grateful to Shunsuke Yabunaka for his critical reading of the manuscript and for valuable comments.
The author acknowledges financial support from JST FOREST (Grant No. JPMJFR252A).
The author used OpenAI's ChatGPT for assistance with language editing, and preparation of figures and numerical scripts.
\end{acknowledgments}

\section*{DATA AVAILABILITY}
The data that support the findings of this study are available from the corresponding author upon reasonable request.

\bibliography{electrochem}
\end{document}